\documentstyle[psfig]{l-aa}
\newcommand{\etal}{et al. }
\newcommand{\kms}{km~s$^{-1}~$}
\newcommand{\msol}{M$_{\odot}$~}
\newcommand{\rsol}{R$_{\odot}$~}

\newcommand{\pyr}{yr$^{-1}$}
\begin{document}
\thesaurus{07                 % Stellar atmospheres
          (03.20.4;           % Techniques: photometry
           03.20.7;           % Techniques: radial velocities
           08.02.1;           % Binaries: close
           08.09.2:HD~101584; % Stars: HD~101584
           08.13.2;           % Stars: mass-loss
           08.16.4)}          % Stars: AGB and post-AGB
\title{The 218~day  period of the peculiar late B-type star
HD~101584\thanks{Based on observations collected at ESOand the
Long-Term Photometry of Variables project}}
\label{ch-art5}
\author{Eric J. Bakker\inst{1,2}\and
        Henny J.G.L.M. Lamers\inst{2,1}\and
        L.B.F.M. Waters\inst{3,4}\and
        Christoffel Waelkens\inst{5}}
\offprints{Eric J. Bakker, present address:
           Astronomy Department,
           University of Texas,
           Austin, TX 78712, USA,
           ebakker@astro.as.utexas.edu}
\institute{Astronomical Institute, University of Utrecht, P.O. Box 80.000,
           NL-3508 TA Utrecht, The Netherlands
\and
           SRON Laboratory for Space Research Utrecht,
           Sorbonnelaan 2, NL-3584 CA Utrecht, The Netherlands
\and
           Astronomical Institute, University of Amsterdam,
           Kruislaan 403, NL-1098 SJ Amsterdam, The Netherlands
\and
           SRON Laboratory for Space Research Groningen,
           P.O. Box 800, NL-9700 AV Groningen, The Netherlands
\and
           Astronomical Institute, Katholieke Universiteit Leuven,
           Celestijnenlaan 200, B-303 Heverlee, Belgium}
\date{Received May 1995, November 1995}
\maketitle

\begin{abstract}
We have searched for periodicity in the photometric and spectroscopic
variations of the enigmatic star HD~101584 and found a long-term
variability on a typical time scale of 1700~days (4.7~years) and
a highly significant period of $218\pm0.7$~day.
This period is most prominently present
in the photometric indices which are a measure for the Balmer jump
(Geneva $d$ and Str\"{o}mgren $c_{1}^{0}$).
The Doppler velocities
of the high-excitation photospheric absorption lines (He\,{\sc I}
and C\,{\sc II}) seem to
be variable with the photometric period. Our data favor
the 218~day period for the Doppler velocities with a small
probability that the true period is 436~days.  We argue that HD~101584
is a close (highly) eccentric 218~day
binary system with a low-mass unseen secondary.
The photometric and Doppler variations are
attributed to changes in the velocity law and mass-loss rate
of the stellar wind which
lead to asymmetric line profiles and a phase dependent Balmer discontinuity.
Binary interaction is responsible for the changes in
velocity law and mass-loss rate leading to the observed phenomena.

\end{abstract}

\section{Introduction}

Since Humpreys (\cite{art5humpreys}), on the basis of limited
data,  suggested
that the peculiar supergiant HD~101584 is
a binary system with an orbital period of 3.5~years,
it has been an open question
whether the binary nature could be proven independently. The last two decades
no convincing improvements have been made in detecting a periodicity.
Here we will show that the optical photometry shows a highly significant
periodicity of $218.0\pm0.7$~days and that the same period is probably
present in the Doppler velocities of high-excitation absorption lines.

Parthasarathy and Pottasch (\cite{art5parthasarathy}) and
Trams~\etal (\cite{art5tramsetal91})
used the detection of infrared excess in the IRAS bands as an argument for
the evolved nature of HD~101584. They interpreted this as
radiation from the AGB ejecta and argue that HD~101584
is in the post-AGB phase.  The galactic latitude of $6^{o}$ is too low
to decide between population I and II, but the observed
radial velocity of HD~101584, derived from the
CO millimeter line emission
($v_{\ast CO}=50$~\kms, Trams~\etal \cite{art5tramsetal90})
cannot be fitted with the Galactic rotation curve if the star is
a massive population I object. This strongly points
to the population II nature of the object.

Bakker~\etal (\cite{art5bakkerart2}) have made an extensive study
on the nature of HD~101584 by combining multi-wavelength
photometry and spectroscopy and found that the system contains a
late B-type post-AGB stars which experiences a very high mass-loss
rate ($\dot M \simeq 10^{-8}$~\msol~\pyr). Due to the high density, the
wind has a low degree of ionization,
close to the star, the ionization conditions in the wind are similar
to those in the photosphere of a F-type supergiant. This is observed
as a low ionization shell spectrum.
Due to the numerous UV spectral lines in the shell
spectrum the UV continuum is not observed but obscured by a curtain
of absorption lines (Bakker~\etal \cite{art5bakker94}).
In the red spectrum the shell spectrum produces
less spectral lines and the late B-type star is observed by the presence
of high-excitation absorption lines (e.g., He\,{\sc I} and C\,{\sc II}).

The most obvious way to look for binarity is to study the Doppler velocities
of spectral lines, which gives the orbital period and the velocity
amplitude of the observed star. This method cannot easily be applied to
HD~101584 because of the velocity-stratification in the
pseudo-photosphere. The Doppler velocities of
absorption lines are not only dependent on the orbital phase, but also
on the wind structure, line strength and central wavelength of the line
(Bakker \cite{art5bakker93}; Bakker~\etal \cite{art5bakkerart2}).
A radial velocity curve on the basis of
high-excitation photospheric absorption lines may prove the binarity.

We have looked for periodicity in the photometric data of HD~101584.
One of us, CW, has observed HD~101584 over many years,
and obtained 89 observations in the Geneva system. The
second, independent, dataset we have used is from the Long-Term
Photometry of Variables (LTPV) project. These 75  observations have been
obtained in the Str\"{o}mgren system, were the $y$-band has been transformed
to the  Johnson $V$-band ($V_{J}$).
A description of the observations is given
in Sect.~\ref{art5secobs}.
We have combined the Geneva $V$-band with
the LTPV $y$-band to obtain 157 Johnson $V_{J}$ magnitudes over 3337 days
(4.1~years). Using the Phase Dispersion Minimization method
(Sect.~\ref{art5secpdm}), we have detected a $218.0\pm0.7$~day
periodicity in $V_{J}$ (Sect.~\ref{art5secphot}). In
Sect.~\ref{art5secdop} we will show that the 218~day period is likely
present in the Doppler velocities of high-excitation photospheric
absorption lines,
and discuss the results in Sect.~\ref{art5secdis}.

\section{The observations}
\label{art5secobs}
\subsection{Photometry in Geneva seven color system}

CW has collected photometry in the Geneva seven color system
($U$, $B1$, $B$, $B2$, $V1$, $V$, $G$)
spanning 1511 days (4.1~years).
These observation were obtained with the 70cm Swiss telescope
at ESO, La Silla, Chile.
The error estimate on the $V$ magnitude is 0.003~mag.
A complete description of the Geneva seven color system can be found in
Rufener and Nicolet (\cite{art5rufenernicolet}).
We will use the Geneva photometric indices $B2-V1$, $d$ and $g$ which
are a measure of the effective temperature, the Balmer
discontinuity, and the blocking by absorption lines respectively.

\subsection{Photometry in Str\"{o}mgren $uvby$ system}

We have obtained  data from the Long-Term Photometry of Variables (LTPV)
project (Sterken \cite{art5sterken1983}) covering
10.6~years.  All observations were taken using
the Danish 50cm telescope on La Silla, and are made in the Str\"{o}mgren $uvby$
system. The photometry of the $y$-band has been transformed to the Johnson
$V_{J}$-band using the relation between those two bands
(Manfroid~\etal \cite{art5manfroidetal}). The data has been published by
Manfroid~\etal (\cite{art5manfroidetal}) and
Sterken~\etal (\cite{art5sterken1993}).
We will use the Str\"{o}mgren indices $(b-y)_{o}$, $c_{1}^{0}$,
and $m_{1}^{0}$, which
are a measure of the effective temperature, Balmer jump,
and blocking by absorption lines respectively.

We have calibrated the Str\"{o}mgren photometry of HD~101584
using the method described by
Spoon~\etal (\cite{art5spoonetal}).
The LTPV program makes use of comparison stars
to increase the accuracy of the data. Just before and just after
observing the program star, comparison stars  are observed.
Comparison star A is HD~102113 (A0IV) and comparison star B is
HD~101735 (G5III).
By assuming that the two comparison stars
are intrinsically constant in magnitude we can calculate a
correction term by which the observed magnitude of the comparison
stars differs from their average value over the whole dataset.
The observed magnitude of the program stars is corrected with
this correction term to obtain an ``atmospheric extinction'' free
intrinsic
magnitude (Eq.~\ref{art5eq-calib}). In order to correct the data,
the comparison stars have to be observed
within 30 minutes before or after the program star. Observations
of the program star within  three minutes of each other were averaged
to one magnitude.

\begin{equation}
\label{art5eq-calib}
m_{corrected} = m_{observed} - \frac{m_{A}+m_{B}}{2}
                            +  \frac{< m_{A}+m_{B} >}{2}
\end{equation}

\noindent
where \( <m_{A}+m_{B}> \) is the mean magnitude of stars A and B
averaged over all observations.

\subsection{The Johnson $V_{J}$-band}

The filter response function of the Geneva $V$-band and the Str\"{o}mgren
$y$-band, as converted to the Johnson $V_{J}$-band,
are almost identical. For this work we
have made use of this and compiled a dataset of Johnson $V_{J}$-band photometry
containing the Geneva $V$-band
and the calibrated Str\"{o}mgren $y$-band. This dataset contains 157
data points and will be referred to as the Johnson $V_{J}$-band
dataset  throughout this article.

\subsection{Doppler velocities}

High-resolution CAT/CES spectra (Table~\ref{art5tab-velcurve}, 17 spectra)
have been obtained by CW in
the wavelength region of He\,{\sc I} (5876~\AA) and C\,{\sc II}
(6579 and 6583~\AA).
Although there are many lines in the optical spectrum of HD~101584,
the majority is formed in the stellar wind
(Bakker~\etal \cite{art5bakkerart2})
and only the high-excitation absorption lines (e.g., He\,{\sc I}
and C\,{\sc II})
are from the photosphere of the underlying
B-type star. Here we will try to correlate the photometric
period with the Doppler velocities  of these high-excitation absorption lines.
All He\,{\sc I} lines were fitted with a Gaussian to derive
the central velocity. The C\,{\sc II} lines are too
weak for a reliable Gaussian fit and the central velocity was determined
with an eyeball fit and the equivalent width by integrating over the
line profile. If more than one velocity is available at a given date,
the velocity used for our analysis was based on the quality of the spectra
(Table~\ref{art5tab-velcurve}). A typical error in the velocity is 3~\kms.
All velocities are heliocentrically corrected.

\begin{table*}
\caption{Heliocentric radial velocities of the C\,{\sc II}
(6579 and 6583~\AA)  and He\,{\sc I} (5876~\AA)  absorption lines.}
\label{art5tab-velcurve}
\centerline{\begin{tabular}{llllrlll}
\hline
Date    &  $JD$ (244+)&Id.& $\lambda_{obs}$ & $W$  & $v_{\rm rad}$ &
Remark & $\overline{v_{JD}}$ \\
dd/mm/yy&     & & [\AA]                 & [m\AA]& [\kms]&
       & [\kms]    \\
\hline
        &        &    &       &    &   &             &       \\
17/02/89& 7575.85& C\,{\sc II}&6578.69&  35& 48&             &  48.0 \\
        &        & C\,{\sc II}&6583.55&  25& 48&             &       \\
\cline{8-8}
23/01/90& 7915.43& C\,{\sc II}&6578.87&  38& 57&             &  57.3 \\
        &        & C\,{\sc II}&6583.45&  14& 46& very noisy  &       \\
\cline{8-8}
02/07/90& 8074.75& He\,{\sc I}&5876.98& 111& 52&             &  51.8 \\
\cline{8-8}
14/03/91& 8329.65& C\,{\sc II}&6579.01&  30& 55& very noisy  &       \\
        &        & C\,{\sc II}&6583.61&  24& 45& very noisy  &       \\
\cline{8-8}
29/01/92& 8650.86& C\,{\sc II}&6578.61&  ? & 45& very noisy  &       \\
        &        & C\,{\sc II}&6583.52&  19& 49& noisy       &       \\
\cline{8-8}
17/04/92& 8729.50& C\,{\sc II}&6579.18&  44& 53&             &  52.6 \\
        &        & C\,{\sc II}&6583.92&  26& 49&             &       \\
\cline{8-8}
18/01/93& 9006.0 & He\,{\sc I}&5876.32& 158& 55&             &  55.3 \\
18/01/93& 9006.0 & C\,{\sc II}&6578.81&  48& 55&             &  52.4 \\
        &        & C\,{\sc II}&6583.52&  53& 50&             &       \\
\cline{8-8}
11/02/93& 9030.0 & C\,{\sc II}&6579.05&  86& 64&             &  64.1 \\
        &        & C\,{\sc II}&6583.77&  45& 59&             &       \\
\cline{8-8}
12/02/93& 9031.0 & He\,{\sc I}&5876.49& 143& 62&             &  61.8 \\
\cline{8-8}
22/04/93& 9089.69& He\,{\sc I}&5876.81& 155& 59& asymmetric  &  59.2 \\
22/04/93& 9098.95& C\,{\sc II}&6579.30&  48& 56&             &  56.3 \\
        &        & C\,{\sc II}&6584.12&  29& 56&             &       \\
\cline{8-8}
14/05/93& 9121.54& C\,{\sc II}&6579.52&  33& 60&             &  58.5 \\
        &        & C\,{\sc II}&6584.27&  31& 56&             &       \\
15/05/93& 9123.46& He\,{\sc I}&5876.90& 162& 57&             &  57.3 \\
\cline{8-8}
19/07/93& 9188.46& He\,{\sc I}&5877.10& 145& 57&             &  57.0 \\
20/07/93& 9189.50& C\,{\sc II}&6579.69&  43& 57& nice profile&  57.9 \\
        &        & C\,{\sc II}&6584.55&  35& 59& nice profile&       \\
\cline{8-8}
14/12/93& 9335.83& C\,{\sc II}&6578.89&  31& 55& nice profile&  55.3 \\
        &        & C\,{\sc II}&6583.61&  22& 51& polluted    &       \\
        &        &            &    &   &             &       \\
\hline
\end{tabular}}
\end{table*}

\section{Phase Dispersion Minimization method}
\label{art5secpdm}

Searching for periodicity with Fourier analysis techniques one
looses information in trying to fit a sinus shaped curve to the mean
light curve. This loss of information does not occur using
the Phase Dispersion Minimization method (PDM).
The  PDM method was first proposed by Lafler and
Kinman (\cite{art5lafler}) for finding the period
of a RR-Lyrae
variable. Stellingwerf (\cite{art5stellingwerf}) made a generalization of this
PDM method and applied it to the double-mode Cepheid BK Centauri.
This method looks for
periodicity in the light curve by means of an
auto-correlation method that does not require the light curve to have any
pre-known shape, but only that it repeats itself every period.

A set of $N$ photometric observations contains $N$ pairs of
Julian dates $JD$ and observed magnitudes: ($JD_{i},V_{i}:i=1,N$).
Let the variance of the magnitude, $\sigma^{2}$, be given by

\begin{equation}
\label{art5eq-variance}
\sigma^{2} = \frac
                 {\sum_{i=1}^{i=N} \left( V_{i}-\overline{V}\right)^{2}}
                 {N-1}
\end{equation}

\noindent
where $\overline{V}$ is the average magnitude
\( \overline{V}=\sum V_{i} /N \).

The dataset is folded over a given period $\Pi$,
to determine the phase
of the observation dates $\phi= \left( JD - JD_{0} \right) / \Pi ~mod~ 1$.
Dividing the phase in $N_{b}$
bins and covering every bin $N_{c}$ time with a bin-offset of
$1/(N_{b}N_{c})$ we have $M=N_{b}N_{c}$
{\sl dependent} samples of the mean light curve.

For each $j$-th sample containing $M$ points
the variance $s_{j}$ (Eq.~\ref{art5eq-variance-sj})
can be computed analog
to Eq.~\ref{art5eq-variance}.

\begin{equation}
\label{art5eq-variance-sj}
s_{j}^{2} = \frac
             {\sum_{k=1}^{k=n_{j}} \left( V_{k}-\overline{V_{j}}\right)^{2}}
             {n_{j}-1}
\end{equation}

\noindent
The overall variance for the given period $\Pi$ is then given by

\begin{equation}
\label{art5eq-variance-m}
s^{2} = \frac
         {\sum_{j=1}^{j=M} \left( n_{j}-1 \right) s_{j}^{2}}
         {\sum_{j=1}^{j=M} \left( n_{j}-1 \right)          }
\end{equation}

\noindent
we use the ratio

\begin{equation}
\label{art5eq-theta}
\Theta=\frac{s^{2}}{\sigma^{2}}
\end{equation}

\noindent
If the phase diagram shows a correlation
for a given period $\Pi$, the overall variance $s^2$ will be
smaller than the dataset variance $\sigma^{2}$.
This gives
a minimum in the $\Theta$-statistics (Eq.~\ref{art5eq-theta}).
If the phase diagram does not
show any correlation but only a random distribution of magnitudes, the
overall variance $s^{2}$ will equal the dataset variance $\sigma^{2}$, and
$\Theta$ will be close to 1.

The strength of the PDM method is that no pre-known shape of the light curve
has to be given. For the photometric data we
taken $(N_{b},N_{c})=(10,2)$.
This means that we split the light curve in 10 bins, each covering 0.1
part of the photometric phase. Each data point will be used twice.
The $\Theta$-statistics was
calculated
for 0.0 to 0.01~days~$^{-1}$ in frequency steps of
$1 \times 10^{-5}$~days$^{-1}$.

\section{Photometry}
\label{art5secphot}

\subsection{The long-term photometric variability with a time scale of
4.7~years}

\begin{figure*}
\centerline{\hbox{\psfig{figure=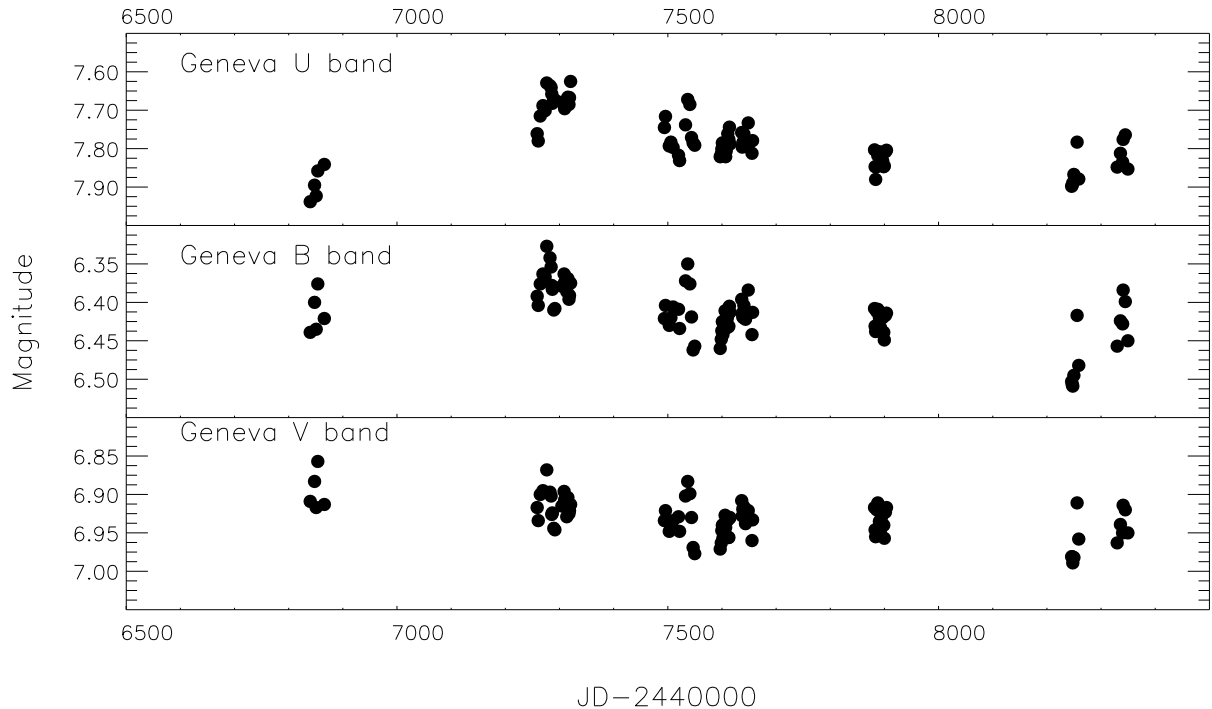,width=\textwidth,rheight=10.0cm}}}
\caption{The magnitudes of the Geneva broad band filters as function
of Julian date show that there is a long-term variation in the
photometry on a typical time scale of 1700~days (4.7~years).}
\label{art5fig-long}
\end{figure*}

The photometric data of HD~101584 shows a long-term variation on
a typical time scale of 1700~days (4.7~years).
Fig.~\ref{art5fig-long} shows
the $U$, $B$, and $V$ broad band filters of the
Geneva dataset (until $JD=2448349.6$). While the $U$-band has a minimum near
$JD=7700$, the $V$-band has a maximum. In the next sections we will discuss
the 218~day photometric period and find a similar behavior which
is attributed to changes in the wind structure and successively in
the Balmer jump.
It is suggestive to think that this variability is periodic, but
our dataset is to limited to reach a firm conclusion about that.
We note that the Kelvin-Helmholtz time scale for a post-AGB star
is of the order of several decades. This might
indicate that the origin of the variability has to do with the
release of gravitational energy of a part (envelope ?) of the star itself.
However the primary goal of this study it to look for photometric
variations which might indicate that HD~101584 is an interacting
binary system. It is unlikely that the 4.7~year variability is
due to binary motion as in such a wide binary system the mutual
interaction will be very small.
Our main goal is to understand the large scatter around
this long-term variation: the 218~days period.

\subsection{The $218.0\pm0.7$~day periodicity in the Johnson $V_{J}$-band}

In order to look for a short-time periodicity on a long-term
variability we should first try to correct for the long-term
variability. However the shape of the intrinsic long-term variation is
unknown, and we have tried to correct for this long-term variation
without satisfactory results. The reason for this is that the PDM
method is very suitable for finding a short-term variation on top of
a long-term variation. By correcting the photometry we
would pollute the light-curve with and ``arbitrarily'' long-term
light curve without gaining much in sensitivity of detection a short
time periodicity.

\begin{figure*}
\centerline{\hbox{\psfig{figure=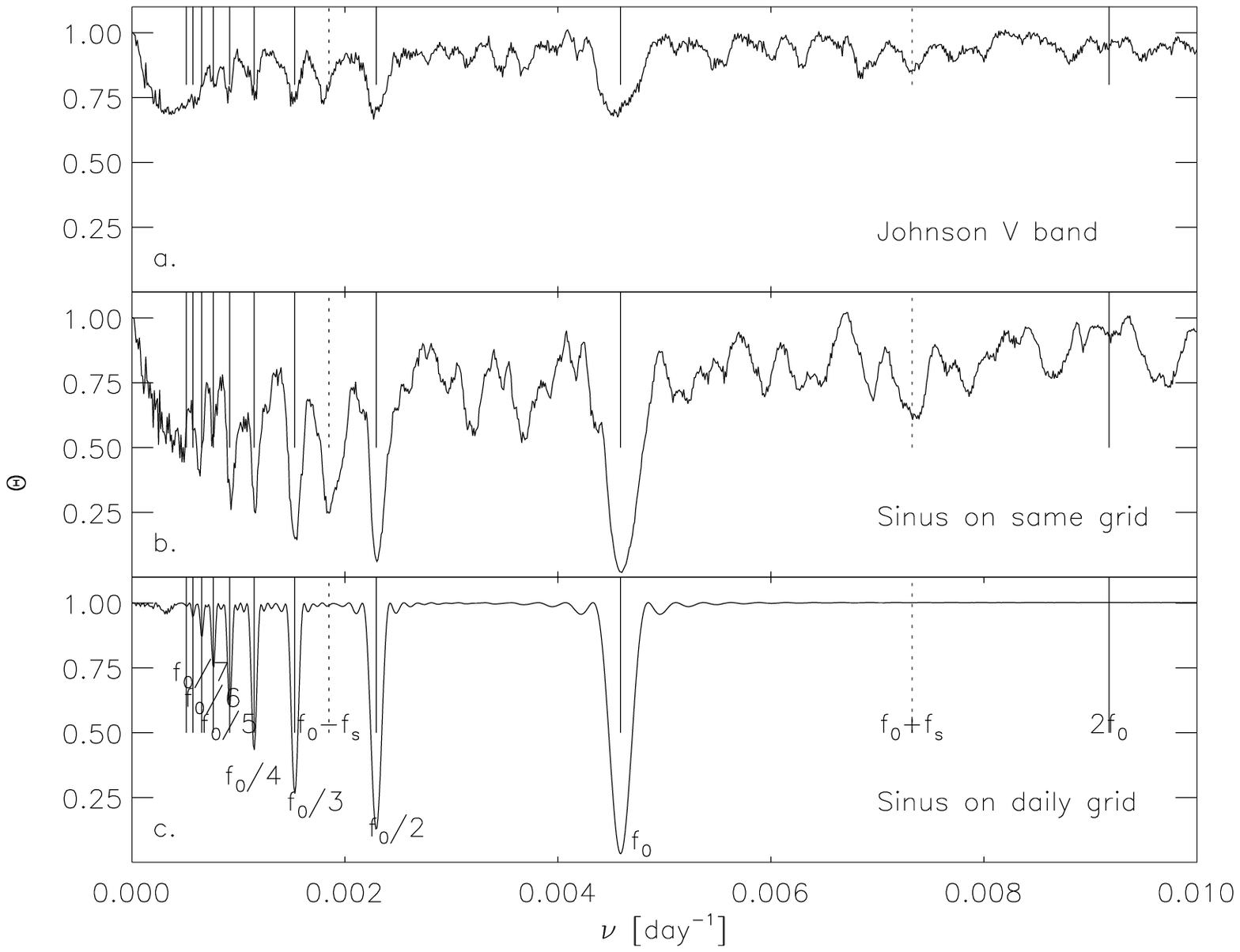,width=\textwidth}}}
\caption{a) $\Theta$-statistics on the 157 data points of the Johnson
$V_{J}$-band.
At least five sub-harmonics can be identified (solid lines).
On both sides of the main frequency
the beat with the year is present (dashed line). b) $\Theta$-statistics on
the grid of
observed dates with a synthetic sinusoidal light curve best representing the
observations. c) the same as ``b'',
but on a daily grid (see text for explanation).}
\label{art5fig-thetaph}
\end{figure*}

Using the PDM method as described in the preceding section we
find a $\Pi=218.0\pm0.7$~day period in the Johnson $V_{J}$-band of HD~101584
(Fig.~\ref{art5fig-thetaph}).
The $\Theta$-spectrum does not only show the primary frequency,
$f_{0}=456.446\times 10^{-5}~{\rm days^{-1}}$,
but also at least  five sub-harmonics and
two beats with the year. This is a highly significant period in the
Johnson $V_{J}$-band.  Table~\ref{art5tab-period} gives a list of the
identified
features in the $\Theta$-spectrum.
Using this period we can make a least-square-fit
to the date using a sine function (Table~\ref{art5tab-sine}).

The many unidentified features in the $\Theta$-spectrum are likely
due to aliases as a result of the irregular spaced grid of observation
dates. We have made a synthetic light curve using a sine-function and
the photometric parameters of Table~\ref{art5tab-sine}.
Such a synthetic light curve on the same grid
of observation dates should show the same features in the
$\Theta$-spectrum. If this is not the case then this could
mean that there is another period present in the data.
Fig.~\ref{art5fig-thetaph}.b shows the $\Theta$-statistics
of the best-fitted-sine-function (Table~\ref{art5tab-sine})
on the grid of observation dates. All features are present
but the power in the $\Theta$-spectrum is much more concentrated
in the (sub-) harmonics of the sine function. This tells use that
it should be possible to obtain better results using another
shape of the synthetic light curve (e.g., sawtooth),
or subtracting the long-term variability.

Finally we have computed a synthetic spectrum on a daily grid between
the first and last day of observations present in the data set.
The resulting $\Theta$-spectrum
(Fig.~\ref{art5fig-thetaph}.c) shows very strong features
at the primary frequency and at least seven sub-harmonics, no harmonics.
The beats with the year are no longer present.
The peak at primary frequency shows lobes on both sides
corresponding to a frequency of the total time span of observations.

A 218~day period in the $\Theta$-spectrum does not unambiguously means
that the true period is 218~days. In the case the photometric
periodicity is due to the rotating geometry of a binary system the true
period can be twice the photometric period. As long as it is not
clear what causes this periodicity it would be premature to
choose between a 218 and 436~day period.

\begin{figure*}
\centerline{\hbox{\psfig{figure=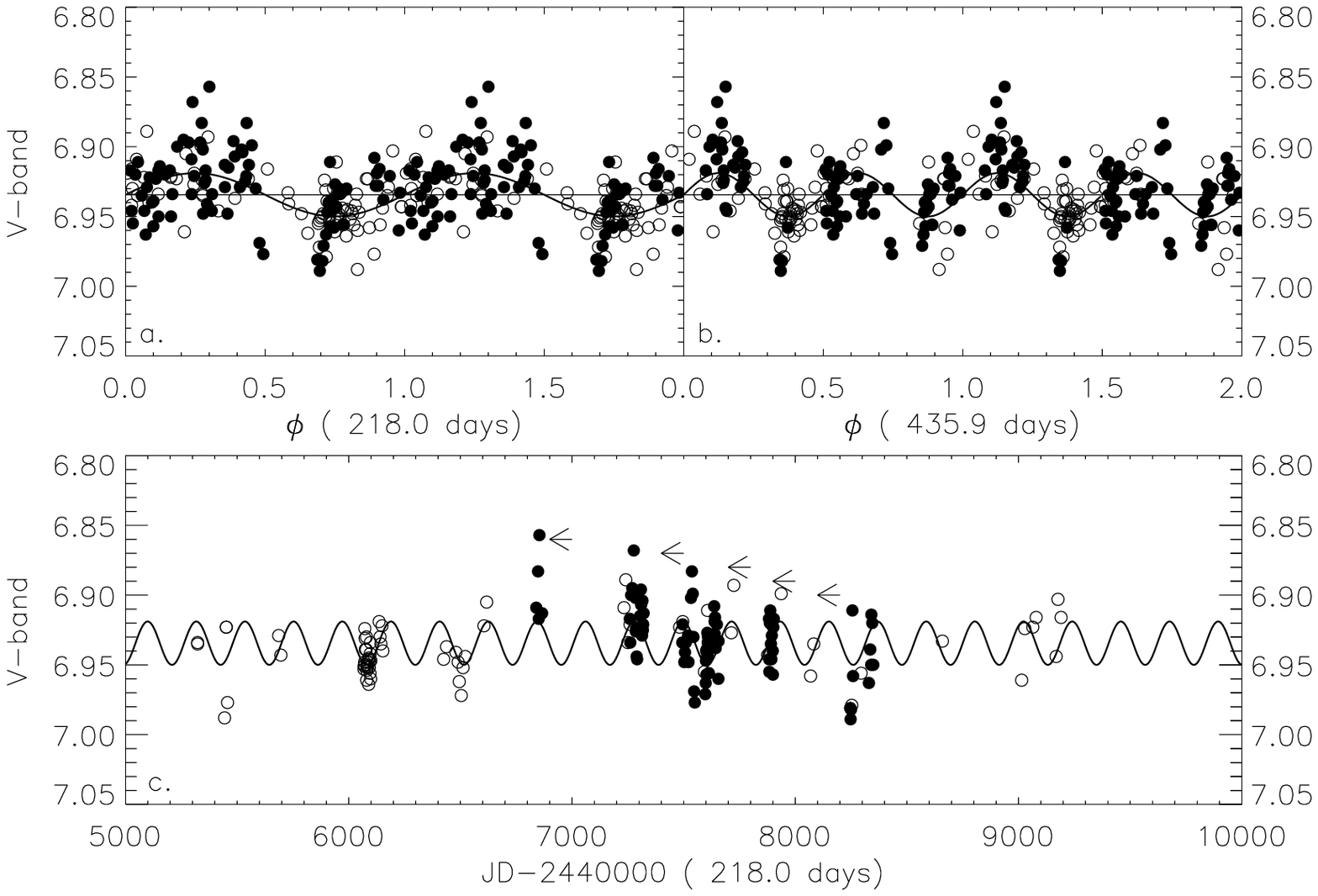,width=\textwidth}}}
\caption{Top two panels the light curve as a function of phase
         (left with $\Pi=218$ and right with $\Pi=436$~days).
         Lower panel the light curve as a function of
         Julian date. In all three cases
         a sine function is over plotted which does best represent
         the data for the period of 218.0~days (this means two
         sines per period for 436~days). Geneva data
         is represented by a dot, and LTPV data by a circle.}
\label{art5fig-lightcurve}
\end{figure*}

Fig.~\ref{art5fig-lightcurve} shows the Johnson $V$ magnitude as function
of phase $\phi$ (based on 218 and 436 day period) and Julian date.
By looking at the light curve as function of Julian date we see that the
magnitude follows the sine function quite well.
We note that there are strong deviations from the mean light curve,
particularly at its maximum.
Unfortunately the photometric light curve
does not significantly show that two successive 218~day
periods behave differently.
The question whether the true period is 218 or 436~days remains therefore
open.

 \begin{table}
 \caption{Identification of the features in the $\Theta$-statistics of
          the Johnson $V_{J}$-band.}
 \label{art5tab-period}
\centerline{\begin{tabular}{lll}
 \hline
 $\nu$   & Identification& Period $\Pi$  \\
 ${\rm [10^{-5}~days^{-1}]}$&& [days]    \\
 \hline
         &                  &            \\
  78.099 & $f_{0}/6$        & 213.40     \\
  91.012 & $f_{0}/5$        & 219.75     \\
 114.783 & $f_{0}/4$        & 217.80     \\
 152.080 & $f_{0}/3$        & 219.18     \\
 181.525 & $f_{0}-f_{s}$    & 219.63     \\
 229.827 & $f_{0}/2$        & 217.55     \\
 456.446 & $f_{0}$          & 219.08     \\
 735.732 & $f_{0}+f_{s}$    & 216.47     \\
         &                  &            \\
 average & $f_{0}$          &$218.0\pm0.7$\\
         &                  &            \\
 \hline
 \end{tabular}}
 \centerline{
 $f_{s}$ is the the sidereal year  of 365.256~days}
 \end{table}

\begin{table}
\caption{Parameters of best fitted sine function.}
\label{art5tab-sine}
\centerline{\begin{tabular}{ll}
\hline
Parameter                & Best fitted value  \\
\hline
                         &                    \\
Period $\Pi$             & $218.0\pm0.7$~days \\
$\overline{V_{J}}$       & $6.934\pm0.008$~mag\\
Amplitude $A_{1}$        & $0.016\pm0.008$~mag\\
$JD(\phi_{\rm phot}=0)$  & $2445044.576$      \\
                         &                    \\
\hline
\end{tabular}}
\centerline{$\phi_{\rm phot}=0$ is defined where
$V_{J}=\overline{V_{J}}$ and $dV_{J}/dt \leq 0$}
\end{table}

\subsection{The $218.0\pm0.7$~day period in the photometric indices}

In order to find the origin of the photometric period we searched
for periodicity in the other photometric bands and photometric indices.
As the dataset is about half the size of the $V_{J}$-band dataset
the PDM method does not give any useful results and we folded
the photometry over a 218~day period and 436~days. In the case of
436~days we find that the two successive sines (of each 218~days)
do not differ significantly. Which means that our data favor a
true period of 218~days. In order to increase the accuracy of
the light curve we will fold the data with a 218~day period.
No clear phase dependence is observed in any of the separate photometric
bands (with the exception of the Johnson $V_{J}$-band because it contains
much more data
points). This is presumably due to the small amplitude and the irregular
behavior of the light curve.
To show the presence of photometric variations we will first look at
the standard deviation ($\sigma$) of the photometry for each
band and index.

\begin{table}
\caption{Statistics on photometry.}
\label{art5tab-stat}
\centerline{\begin{tabular}{|lll|rll|}
\hline
       &    &   &      &      &      \\
band   &$\lambda_{c}$&$\delta \lambda$&$\overline{M}$&
                      $\sigma_{M}$&$\sigma_{M(A/B)}$ \\
\hline
       &    &   &      &      &      \\
$U$    &3458&170& 7.779&0.0710&0.0074\\
$u$    &3500&300& 8.606&0.0684&0.0348\\
$B1$   &4022&171& 7.306&0.0392&0.0054\\
$v$    &4110&190& 7.560&0.0321&0.0172\\
$B$    &4248&283& 6.415&0.0343&0.0048\\
$B2$   &4480&164& 7.890&0.0309&0.0052\\
$b$    &4670&180& 7.262&0.0209&0.0143\\
$V1$   &5408&202& 7.652&0.0250&0.0043\\
$y$    &5470&230& 6.939&0.0192&0.0075\\
$V_{J}$&5500&800& 6.934&0.0228&      \\
$V$    &5508&298& 6.930&0.0249&0.0036\\
$G$    &5814&206& 8.034&0.0243&0.0050\\
       &    &   &      &      &      \\
$B2-V1$&    &   & 0.237&0.0121&0.0033\\
$(b-y)_{0}$&&   &-0.039&0.0077&0.0120\\
$d$    &    &   & 1.308&0.0345&0.0072\\
$c_1^0$&    &   & 0.677&0.0415&0.0390\\
$g$    &    &   &-0.066&0.0091&0.0050\\
$m_1^0$&    &   & 0.037&0.0157&0.0150\\
       &    &   &      &      &      \\
\hline
\end{tabular}}
\end{table}

\begin{figure}
\centerline{\hbox{\psfig{figure=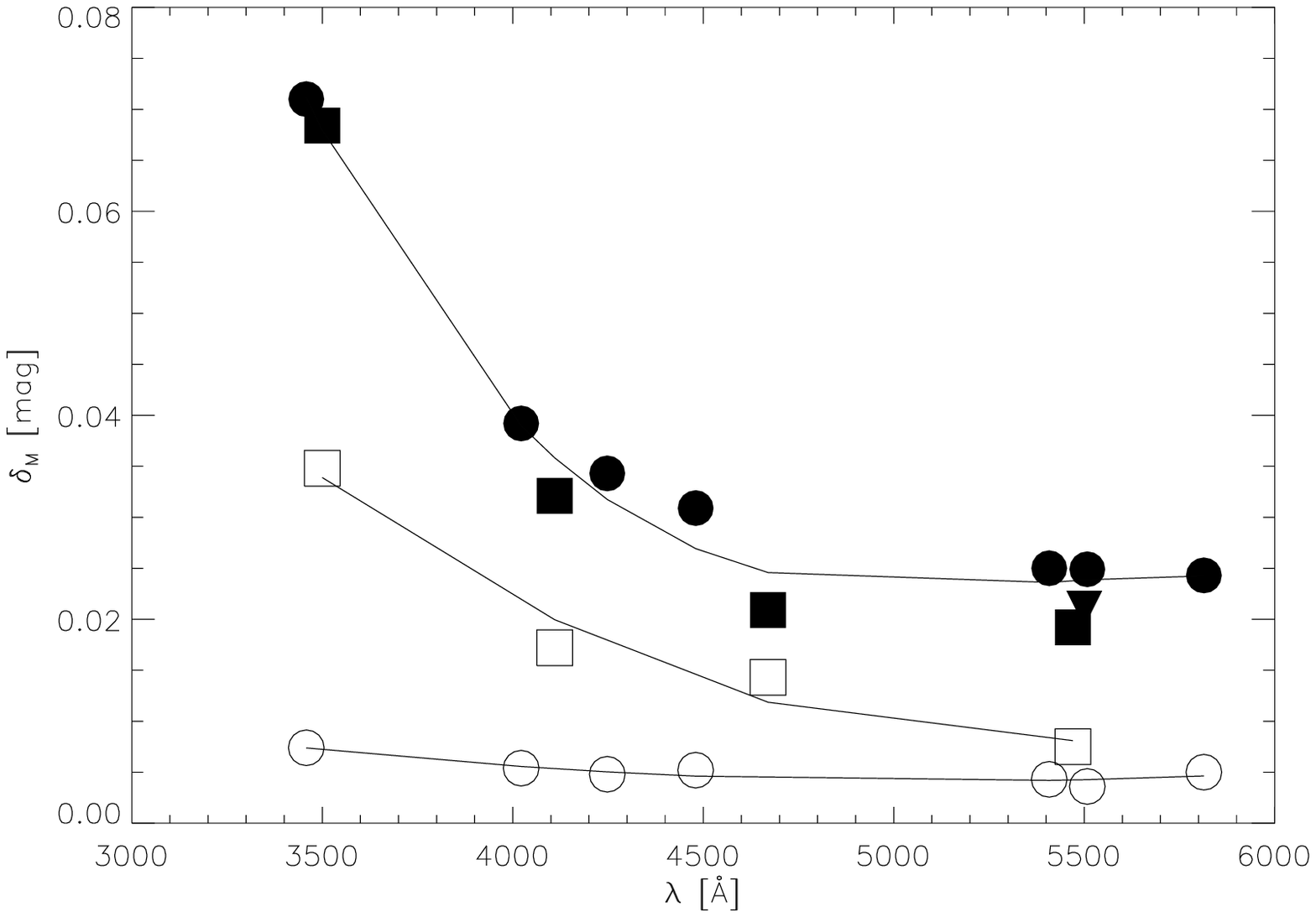,width=\columnwidth}}}
\caption{Standard deviation of the photometric bands ($\sigma_{M}$)
as a function of central wavelength $\lambda_{c}$ of the band.
Clearly the amplitude of the variations increases
to shorter wavelengths and are in excess of the error.
The solid symbols represent the observations of HD~101584
(dots: Geneva, squares: Str\"{o}mgren, triangles: $V_{J}$-band), and
the open symbols the observations of the comparison stars $\sigma_{M(A/B)}$.}
\label{art5fig-amp}
\end{figure}

Table~\ref{art5tab-stat} gives for each photometric band
(in order of increasing wavelength) the
average magnitude ($\overline{M}$) and the standard deviation
($\sigma_{M}$). The last column gives the standard
deviation of that band or index for the comparison stars
($\sigma_{A/B}$) and
is used as a measure of the error in the dataset.
Here we note that the standard deviation on the average  magnitude of
a photometric band or color  is significantly larger than the error.
Fig.~\ref{art5fig-amp}
shows that the standard deviation increases to the blue. Near
the $y$-band the variations are on a 3$\sigma$ level, while due to
an increase of the error the variations decrease
to the blue.
{}From this we conclude that all photometric bands show
variations. Although we can not decide on the basis of Fig.~\ref{art5fig-amp}
whether these variations are correlated with the 218~day period
in the $V_{J}$-band photometry it is suggestive that the 218~day period
is also present in the other photometric bands.

A powerful tool to study the photometric variations are
the photometric indices. Both in the Geneva and in the Str\"{o}mgren
system there are indices for the effective temperature
$T_{\rm eff}$, the Balmer discontinuity, and the blocking by absorption lines.

\begin{figure*}
\centerline{\hbox{\psfig{figure=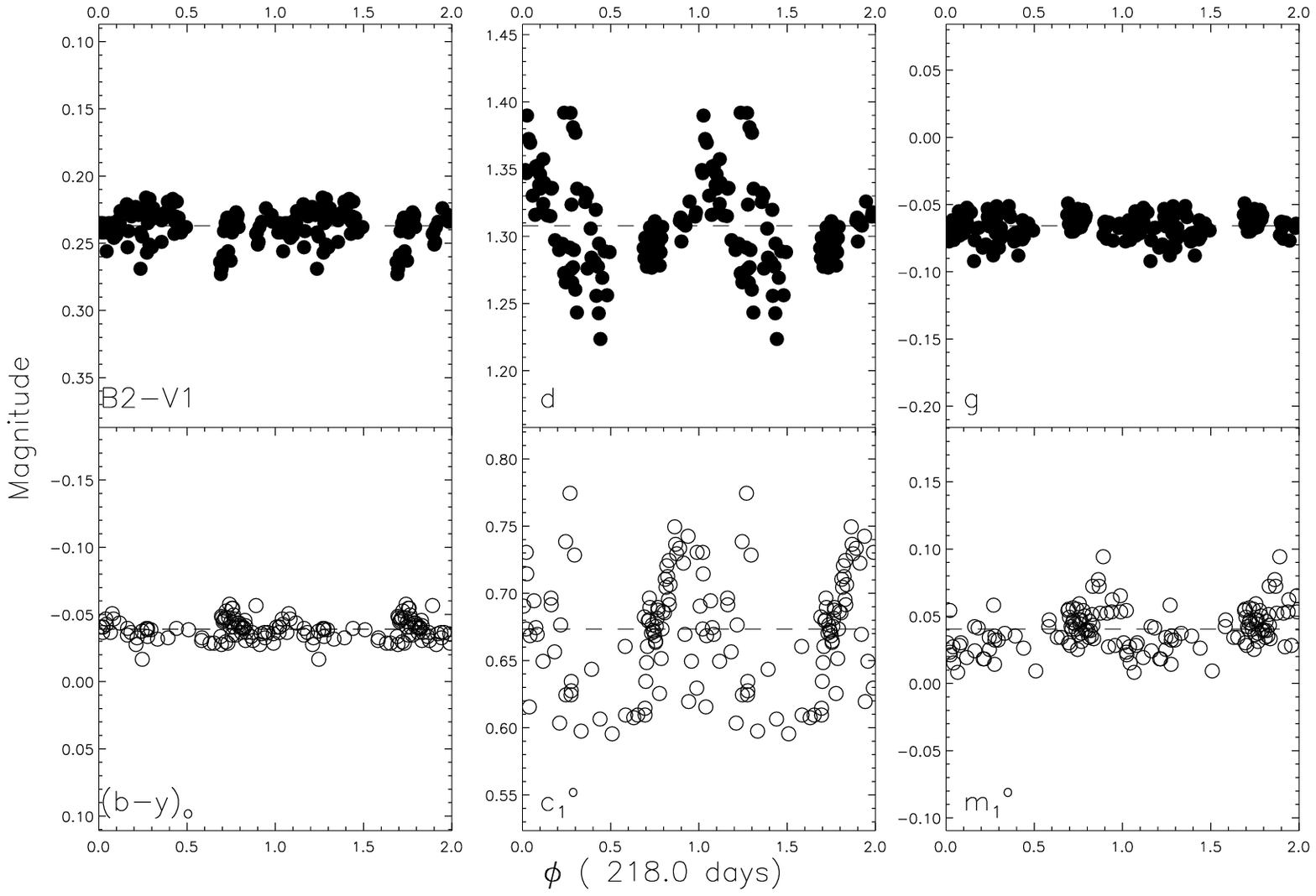,width=\textwidth}}}
\caption{Phase light curve of photometric indices. The phase has
been calculated by folding the data over a 218~day period
and taking $JD=2445044.576$ as zero point. The
three top panel are Geneva indices, while the lower three panels
are Str\"{o}mgren indices (left panels: effective temperature;
middle panels: Balmer discontinuity; right panels: line blanketing).
All six abscissae are scaled to the same
magnitude range of 0.30~mag.}
\label{art5fig-ind}
\end{figure*}

Bakker~\etal (\cite{art5bakkerart2}) have shown that the photometry
is consistent with a late B-type supergiant (e.g., B8-9I-II,
$T_{\rm eff}=12000\pm1000$~K, $\log g=3.0\pm1.0$, and $E(B-V)=0.49\pm0.05$).
To calculate dereddened photometry we will adopt the reddening of
$E(B-V)=0.49\pm0.05$. Using
Eq.~\ref{art5eq-eby} the Str\"{o}mgren photometry can be
dereddened. No dereddening has been applied to the Geneva photometry.

\begin{equation}
\label{art5eq-eby}
E(b-y) = 0.70 \times E(B-V) = 0.343 \pm 0.04
\end{equation}

The effective temperature indicators are $(b-y)_{0}$ (Eq.~\ref{art5eq-b-y})
and $(B2-V1)$ (Eq.~\ref{art5eq-b2-v1}).

\begin{eqnarray}
\label{art5eq-b-y}
\left( b-y \right)_{0} = \left( b -  y \right) - E(b-y) \\
\label{art5eq-b2-v1}
\left( B2 - V1 \right) = B2 - V1
\end{eqnarray}

\noindent
In Fig.~\ref{art5fig-ind}
both indices are folded with the 218~day period and no clear
phase dependence is observed. From this we conclude that the
218~day period is not related to a periodicity in $T_{\rm eff}$.

The photometric indicators for the Balmer discontinuity
are $c_1^0$ (Eq.~\ref{art5eq-c10}) and
$d$ (Eq.~\ref{art5eq-d}).

\begin{eqnarray}
\label{art5eq-c10}
c_1^0=(u-v)-(v-b)-0.20 \times E(b-y)   \\
\label{art5eq-d}
d    =  \left( U - B1 \right) - 1.430 \times \left( B1 - B2 \right)
\end{eqnarray}

\noindent
Fig.~\ref{art5fig-ind} clearly shows the
presence of a 218~day period in those two indices.
The amplitude of the variations in $d$ and $c_1^0$
are about 0.18~mag (a $35 \sigma$ detection), i.e.
a factor eleven larger than the amplitude of
0.016~mag found in the $V_{J}$-band photometry.
The fact that the Geneva $d$ and the Str\"{o}mgren $c_{1}^{0}$
indices are not exactly in phase is likely due to the
fact that they are not observed simultaneously and uses different photometric
bands. Furthermore we should note again that we did not correct for
the long-term photometric variability and that this will surely causes
an additional scatter in the phase light curve.
A maximum Balmer discontinuity
occurs at phase $\phi_{\rm phot}=0.0\pm0.1$, which is clearly out of phase
with the $V_{J}$-band: it is $\delta \phi=-0.25\pm0.10$ behind. The
minimum Balmer discontinuity occurs at phase
$\phi_{\rm phot}=0.5\pm0.1$, exactly
half a period shifted from the maximum Balmer discontinuity.
Suppose that the large standard deviation of the temperature
indicator $(b-y)_{0}$ is due to variations in the effective temperature
of the star. Then the mean relation between $c_1^0$
and $(b-y)_0$, $(b-y)_0=-0.116+0.097 \times c_1^0$,
predicts a standard deviation of the  $c_1^0$ index of
$\sigma=0.0008$. The observed standard deviation is a factor 50 larger
and (again) tells us that it is not the temperature but
the (wind)density that causes the variability of the Balmer discontinuity.
{}From the work by Bakker~\etal (\cite{art5bakkerart2}) we know that
at wavelengths shorter than the Balmer jump (3647~\AA) the wind
spectrum dominates and is optically thick in the Balmer
continuum: the UV energy
distribution resembles an A5I star. The periodicity
in $c_{1}^{0}$ can now be attributed to a periodicity in the
density of wind material (not the photosphere).

Photometric variations due to obscuration by wind material is also observed
in the
case of the well studied object Pleione (Luthardt and Menchenkova
\cite{art5lutmen}).
The spectral type of Pleione of B8V (M=3.5~M$_{\odot}$)
is comparable to the spectral type of HD~101584: B(e)8-9II. In contrast to
HD~101584 with $v \sin i \approx 50\pm10$~\kms (assuming that we don't see
it pole on), Pleione is a rapid rotator ($v \sin i=300$~\kms)
which experiences discrete mass-loss episodes.
A shell of material
is expelled and moves away from the star and the spectrum of a normal B8V star
changes to a spectrum which is almost completely dominated by wind lines which
are partially  in emission.  In the case of Pleione the
recurrence time scale is of the order of several decades, while
the recurrence time scale for HD~101584 is much shorter (218~days)
and the star is almost continuously in a shell stage.

Recent theoretical work on the shell spectrum after a nova outburst
 (Beck~\etal \cite{art5beck}) has shown that in the case of a dense
shell of material around a 10.000~K star, the Str\"{o}mgren sphere
of e.g., hydrogen is one order of magnitude smaller than in the
case of a black-body. This leads to a depression of the UV continuum. The
observed photometric variations of HD~101584 can be understood in
a similar way as a period depression of the UV continuum due to
variations in the amount of gas in the line-of-sight.

The photometric indices for blocking by absorption lines are
$m_1^0$ (Eq.~\ref{art5eq-m10}) and $g$ (Eq.~\ref{art5eq-g}).

\begin{eqnarray}
\label{art5eq-m10}
m_1^0=(v-b)-(b-y)+0.18 \times E(b-y) \\
\label{art5eq-g}
g    = \left( B1 - B2 \right) - 1.357 \times \left( V1 - G \right)
\end{eqnarray}

Fig.~\ref{art5fig-ind} shows that no clear phase dependence of these
indices is observed.  The presence and strength of spectral features
does not vary significantly, which indicates that the periodicity
found can be attributed to a periodicity in the strength of the
UV continuum level.  The number of wind lines in the optical
spectrum of HD~101584 decreases to the red. For
wavelengths longer than 5000~\AA~ there are very few of these lines
left and it unlikely that these lines produce the periodicity in
the photometry.
Furthermore from Fig.~\ref{art5fig-amp} we note
that although the Balmer continuum shows the strongest variability,
there are indications that even the Paschen continuum varies in strength.

\subsection{Conclusions}

We found a long-term variability (possible period) with a
typical time scale of 4.7~years.
The Balmer (and Paschen)
discontinuity varies periodically with
218~days and that the effective temperature and blocking by
absorption lines do not vary. This periodicity is therefore attributed
to changes in the continuum level, and especially in the Balmer continuum
(UV continuum). As the UV continuum is much lower than expected for a
B8 supergiant we argue that the UV continuum is depressed due to the
many absorption lines from a dense stellar wind and that the amount
of depression of the UV continuum is period.

\section{Doppler velocities}

\label{art5secdop}

\subsection{The 218~day period of the Doppler velocities}

An important question in unraveling the photometric periodicity is whether
or not the periodicity is also present in the Doppler velocities of the
photospheric absorption lines, and if it is, whether the period is
218 or 436~days. The small number of data points
(only 15) is too small to detect a period using the PDM technique
and we have folded the observations over 218 and 436~days.

\begin{figure*}
\centerline{\hbox{\psfig{figure=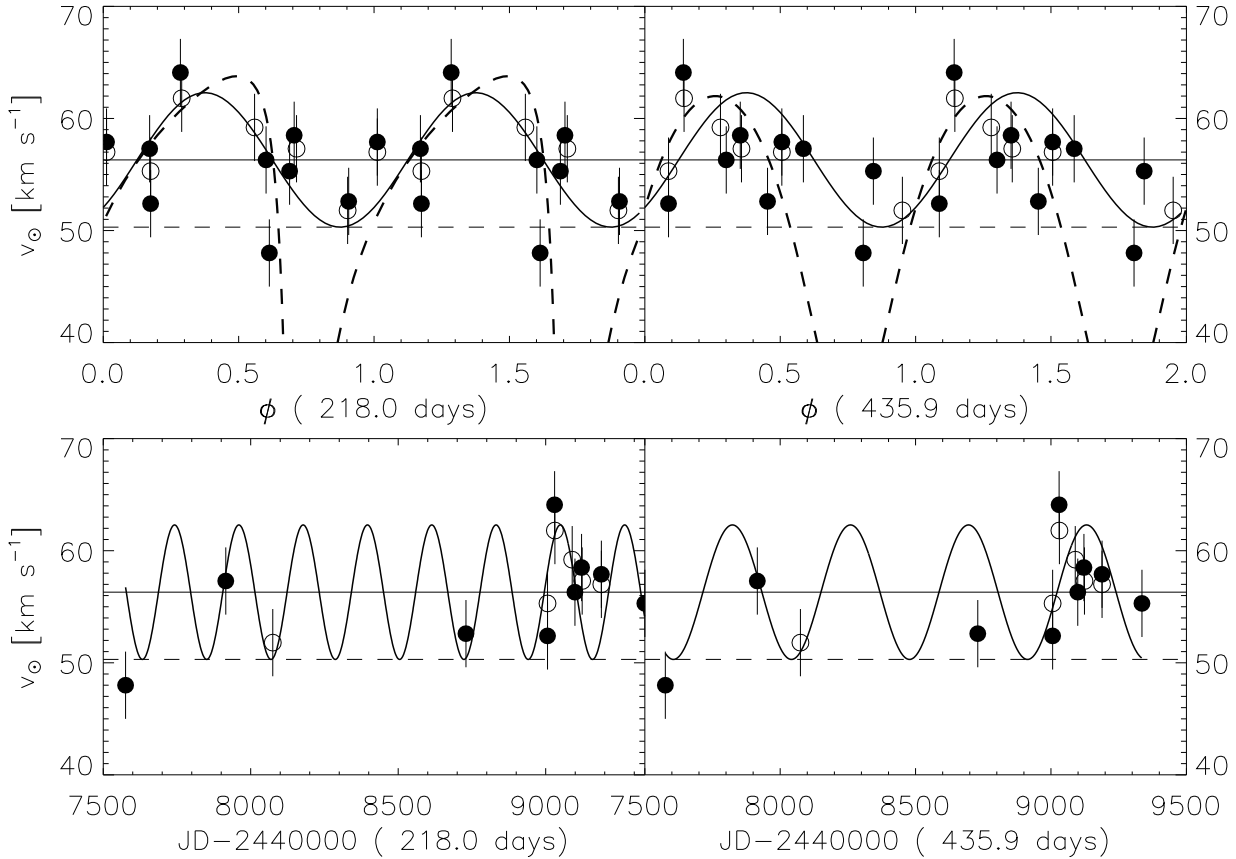,width=\textwidth}}}
\caption{Radial velocity curve of the high-excitation photospheric absorption
lines. The two left panels are made by folding with the photometric period
(218~days),
while the right two panels were folded with twice the photometric
period (436~days). The solid sine is the best fit through the
data points ($e=0$) and the dashed lines are highly eccentric orbits fitted
to the data points(left $e=0.7$ and right $e=0.2$) as a possible explanation
for the 6~\kms velocity offset.
The lower two panels give the radial velocity curve versus Julian
date for  218 and 436~days respectively. C\,{\sc II} is represented by a filled
circle and He\,{\sc I} by an open circle.}
\label{art5fig-velcurve}
\end{figure*}

\begin{table}
\caption{$v_{\rm rad}$: parameters of best fitted sine function.}
\label{art5tab-vel}
\centerline{\begin{tabular}{lll}
\hline
Parameter                    & Best fitted value     \\
\hline
                             &                       \\
Period $\Pi$                 & $218.0\pm0.7$~days    \\
$\overline{v_{\rm System}}$  & $ 56.3\pm1.0$~\kms    \\
Amplitude $K_{1}$            & $  3.0\pm3.4$~\kms    \\
$JD(\phi_{\rm rad}=0)$       & $2444990.08 $         \\
                             &                       \\
\hline
\end{tabular}}
\centerline{$\phi_{\rm rad}$ is defined when
$v_{\rm rad}=\overline{v_{\rm system}}$ and $dv_{\rm rad}/dt \geq 0$}
\end{table}

Fig.~\ref{art5fig-velcurve} shows the two possibilities (218 or 436~days).
On the basis of this figure we are not able to distinguish between
the two periods, but as we have argued that the photometric period
is 218~days we will adopt a 218~days period in the radial velocities.
A sine fit with a period of 218~days yields an amplitude of 3.0~\kms
with a residual dispersion of 3.4~\kms. The fit
is not in phase with $c_{1}^{0}$ but leads it by 0.25 in phase.
It is however very doubtful whether a sine fit is allowed
as the
average velocity of $56.3\pm1.0$~\kms is 6.0~\kms red-shifted with respect
to the system velocity. Most likely due to the small number statistics
the observed average radial velocity is off by about 6.0~\kms. As only
one point is below the system velocity a sine fit with an offset of
the system velocity does not not give a reasonable fit to the data.
An ad hoc interpretation to the 6~\kms offset could be that we don't
see direct light from the star, but only reflected light on a receding
``mirror'' of material (e.g., wind material). As this is indeed an ad hoc
interpretation we will not pursue it.

\begin{figure*}
\centerline{\hbox{\psfig{figure=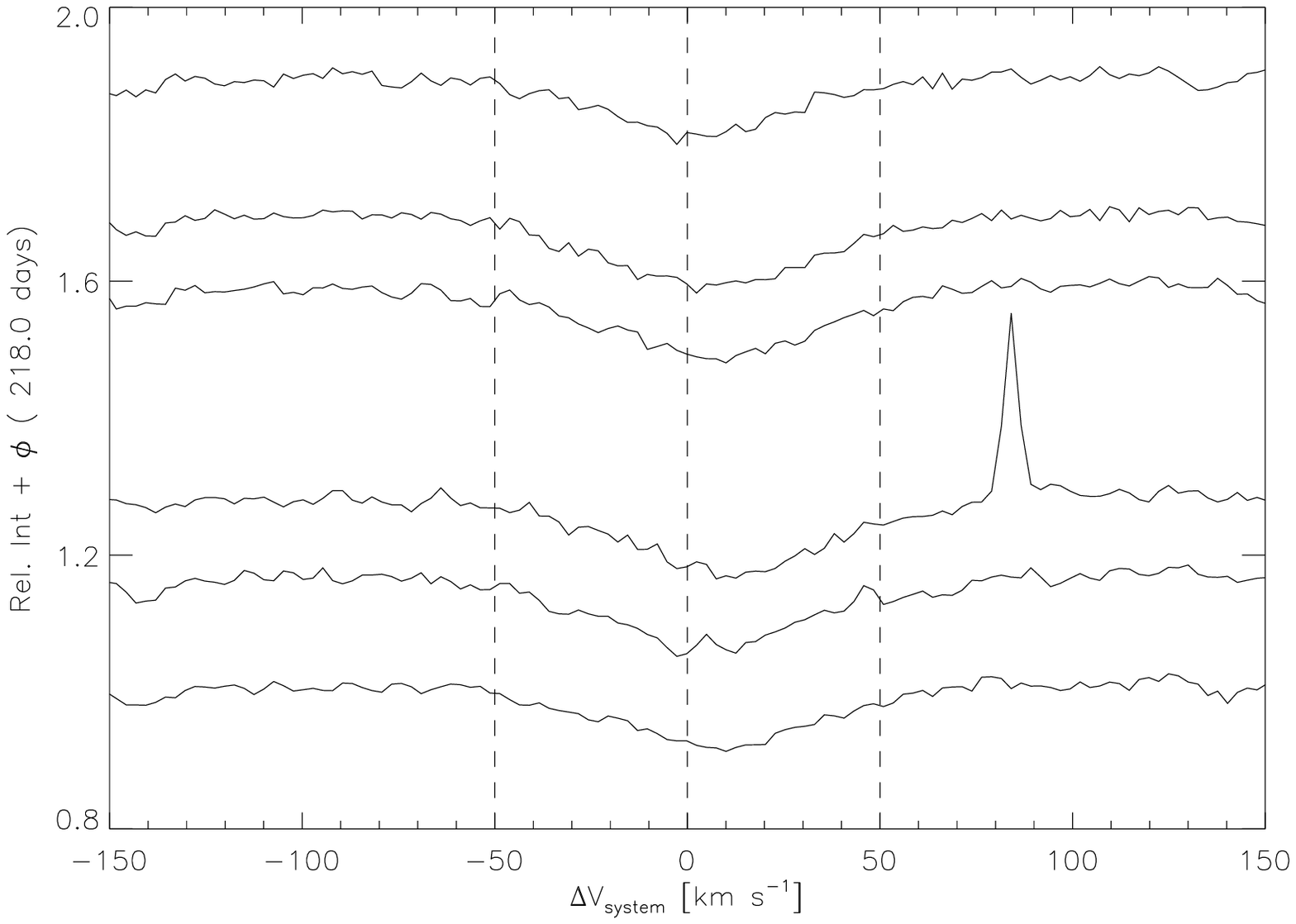,width=\textwidth}}}
\caption{He\,{\sc I} (5876~\AA) lines profiles as function of phase
$\phi_{\rm rad}$.  Each
spectrum is offset with $\phi_{\rm rad}$. This figure demonstrates the
small changes in the position of the center of the profile.}
\label{art5fig-hei}
\end{figure*}

\begin{figure}
\centerline{\hbox{\psfig{figure=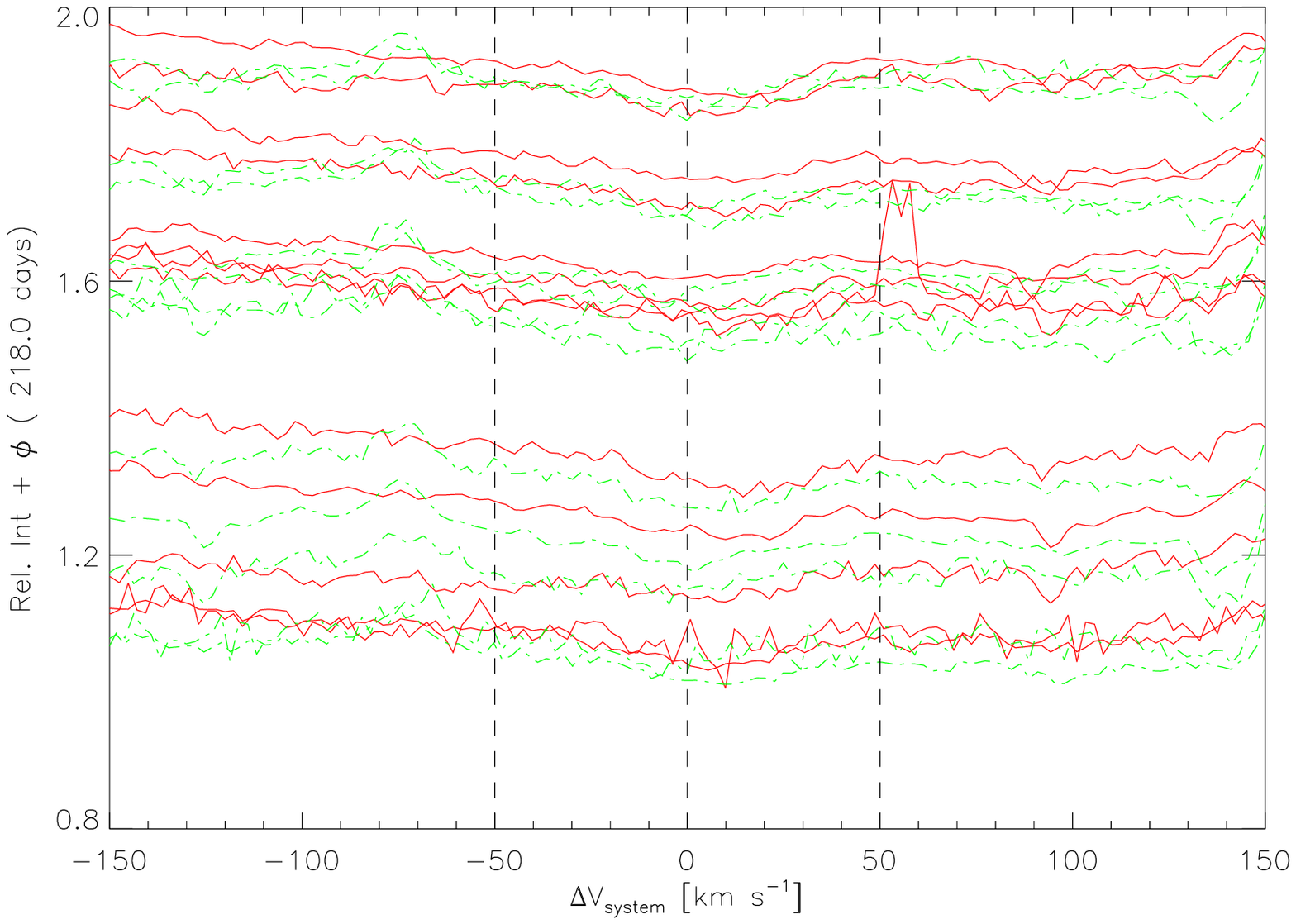,width=\columnwidth}}}
\caption{C\,{\sc II} lines profiles as function of phase $\phi_{\rm rad}$. Each
spectrum is offset with $\phi_{\rm rad}$. This figure demonstrated the
small changes in the position of the center of the profile.
The solid line
is C\,{\sc II} (6578.03~\AA) and the dashed  line C\,{\sc II} (6582.85~\AA).}
\label{art5fig-cii}
\end{figure}

We have investigated if the periodicity in the Doppler velocities
could be attributed to changes in the absorption line profiles.
Fig.~\ref{art5fig-hei} shows the line profile of the He\,{\sc I} lines
at six different phases and Fig.~\ref{art5fig-cii} shows the line profiles
of the two C\,{\sc II} lines at 13 different phases.
The dashed lines is the system velocity of
$50.3\pm2.0$~\kms with on both sides a line at
$v \sin i \approx 50\pm10$~\kms
to show the variable asymmetry of the line profile.
The red wing is steeper
than the blue wing, and there is clearly a shift in the position
of the minimum of the profile. This could cause a systematic error in the
determination of the Doppler velocities.

\subsection{Stellar rotation?}

An estimate of $v \sin i$ can be obtained from the full width
of the absorption line
profile of C\,{\sc II} (Fig.~\ref{art5fig-cii}). The full width of the
He\,{\sc I} line is not a good estimate
because of Stark broadening.
{}From the C\,{\sc II}  lines we find $v \sin i \approx 50\pm10$~\kms.
If 218~days is
the stars rotational period and the system is almost edge-on
($\sin i=1$)
then we find that the radius of
the star is $214\pm50$~R$_{\odot}$.
This is one order of magnitude too large for a
post-AGB star, but is
about the size expected for a massive 40~\msol supergiant.
Alternative if we adopt a
stellar radius ($R_{\ast}=20$~R$_{\odot}$) for a post-AGB star,
and $v \sin i =50\pm10$~\kms
we find a rotational period of $P_{\rm rot}=20$~days.
We can thus conclude that 218~days
is not the rotational period for a post-AGB star
(it could be the rotational period if HD~101584 is a
massive B-supergiant).

\subsection{Pulsation?}

With an amplitude of 6~\kms and a 218~day period the increase of
radius in one pulsation period is 51~R$_{\odot}$.
This is larger than the radius of a post-AGB star ($R_{\ast}=20$~R$_{\odot}$).
Such an enormous
increase in radius
would be observed and strong variations in the photometric indices which are a
measure of the
effective temperature would be observable.
As this is not the case we can discard pulsation as the origin of the
photometric variations.

\subsection{Binarity?}

If the 218~day period in  the Doppler velocities is attributed to orbital
 motion, the mass function of the system
is $f(m)=4.9\times 10^{-3} \left( 1 - e^2 \right) ^{3/2}$~M$_{\odot}$.
{}From Bakker~\etal (\cite{art5bakkerart2}) we take the system
edge-on, $\sin i=1$, and for simplicity
we will adopt $e=0$.

\begin{table}
\caption{Possible orbital parameters
for $f(m)=4.9 \times 10^{-3}$~M$_{\odot}$.}
\label{art5tab-orbit}
\centerline{\begin{tabular}{llll}
\hline
      & $M_1=0.6$~\msol& $M_1=8$~\msol &$M_1=40$~\msol \\
      & $R_1=20$~\rsol & $R_1=24$~\rsol&$R_1=160$~\rsol\\
\hline
      &                &               &               \\
$M_2$ & 0.14~\msol     & 0.72~\msol    & 2.1~\msol     \\
$a \sin i$&$6.9$ R$_{\ast}$&$13$ R$_{\ast}$& $3.3$ R$_{\ast}$\\
$K_2$ & 26~\kms        & 67~\kms       & 116~\kms      \\
      &                &               &               \\
\hline
\end{tabular}}
\end{table}

Table~\ref{art5tab-orbit} gives the orbital parameters for
a 0.6~\msol post-AGB star, a 8~\msol (Pop. I) and a 40~\msol star (Pop I).
The orbital separation, $a \sin i$,
is in all cases less than 13 R$_{\ast}$, and all
systems are close binaries. In all cases the secondary is
a low-mass star, with a mass less than the primary (the observed star).
Stars which are not on the main sequence evolve faster than
on the main sequence. This means that the secondary is most likely
either the end product of stellar evolution, or is on the main sequence.
The secondary of the 0.6~\msol primary is a low-mass main sequence object
as 0.14~\msol white dwarfs do not exist. The secondary of the
8~\msol and 40~\msol primary cannot be a white dwarf as the progenitor was a
high-mass star which evolves to a neutron star
or a black hole which have a mass in excess of 1.4~M$_{\odot}$.
The presence
of a black hole or neutron star is not likely as no high energetic
processes have been reported in the literature.
This secondary is thus also on the main sequence.

A possible solution to the low-mass of the secondary and the velocity offset
of 6~\kms is if the star is in a highly eccentric orbit. This will increase
the amplitude, $K1$, the mass-function, and therefore the mass of the
secondary.
Two examples are plotted in the top two panels of Fig.~\ref{art5fig-velcurve}.
In that case we find post-AGB
secondaries masses in the range between 0.1 and 1.0~M$_{\odot}$.

It is evident that if the variations in Doppler velocities are due to
binarity,
the mass function poses strong constraints on the mass of the secondary
and the eccentricity of the orbit.
The most likely evolutionary status, the post-AGB phase, yields
a mass of the secondary in the range of 0.14~\msol for a
circular orbit to 1.0~\msol in a  (highly) eccentric orbit.

\subsection{Conclusion}

Pulsation and rotation are not able to explain both the photometric
and radial velocity variations.
If the observed star is in a binary system with an eccentric orbit the
radial velocity variations could be due to binary motion which gives
a secondary mass between 0.14 and 1.0~M$_{\odot}$.

\section{Discussion}
\label{art5secdis}

We can summarize the results in the following way: \newline
\begin{enumerate}

\item The photometry reveals a long-term variability on a time scale
      of about 1700~days (4.7~years). On the basis of the available data
      we argue that it is likely periodic.

\item The Johnson $V_{J}$-band, derived from the Geneva $V$
      and Str\"{o}mgren $y$-band,
      shows a highly significant period of $\Pi=218.0\pm0.7$~days.

\item  The period is most clearly present in the photometric indices which
       are a measure of the Balmer jump.

\item  The effective temperature indicators do not vary with phase.

\item  The line blanketing indicators do not vary with phase.

\item The photometric variations are therefore attributed to variations in the
      (UV and optical) continuum level.

\item A change in the strength of  the Balmer jump
      can occur at a part of the stellar surface with
      a lower effective gravity, leading to a higher mass-loss rate
      and a flatter velocity
      law of the wind.

\item The Doppler velocities of high-excitation photospheric
      absorption lines are likely periodic with 218~days, and
      attributed to orbital
      motion in an highly eccentric orbit.

\item The Doppler velocities are not in phase with the Balmer jump.
\end{enumerate}

We have discussed stellar rotation, binarity and pulsation as possible
origin of the 218~day photometric
period and found that only binarity can explain the observations.
(a) HD~101584 is a post-AGB star and the variations are due to
a companion with a period of 218~days.
The changes in the velocity law and mass-loss rate are likely triggered by
binary interacting with the secondary. The asymmetric photospheric
 lines profiles
might cause a systematic effect in the determination of the Doppler
velocities, and
can (partly) explain the 6~\kms offset.
(b) If HD~101584 is a massive object than the stellar rotational period is
the most acceptable origin.

In both scenario the same idea works:
a  part on the stellar surface experiences a flatter velocity law and
higher mass-loss rate  than average.
This results in a periodic increase and decrease of the (UV) continuum level.
The periodicity in $V_{J}$-band is also due to the increase in
the wind density, but as the effect of the amount of neutral hydrogen strongly
decrease to longer wavelength the amplitude is quite small.

A alternative explanation for the variations in the
Balmer discontinuity might be
the presence of a second continuum source in a binary system
(e.g., an accretion disk around the secondary).
The second continuum source is periodically observed or obscured and this
leads to the observed changes in the photometry.

{}From this work we see that the ultraviolet spectral region is of crucial
importance
in understanding
this object as the stellar wind is so dominant. We therefore have started
a long-term
IUE program
to monitor HD~101584 every 40 days at high- and low-resolution.
Here we predict that the UV continuum
will rise and fall in phase with the $c_{1}^{0}$ and $d$ index.

\acknowledgements{EJB was supported by grant
no. 782-371-040 by ASTRON, which receives funds
from the Netherlands Organization for the Advancement of Pure Research (NWO).
LBFMW is supported by the Royal Netherlands Academy of Arts and Sciences.
This research has made use of the Simbad database, operated at CDS, Strasbourg,
France. We acknowledge the Long-Term Photometry of Variables (LTPV) project
for observing this star for such a long period. Without the LTPV data this
article could not have been written. CW thanks the staff of the Geneva
Observatory for the generous awarding of telescope time with the Swiss
Telescope at La Silla Observatory.}

\end{document}